# Mycosubtilin overproduction by *Bacillus subtilis* BBG100 enhances its antagonistic and biocontrol activities

Valerie Leclère[1], Max Béchet[1], Akram Adam[2], Jean-Sébastien Guez[1], Bernard Wathelet[3], Marc Ongena[2], Philippe Thonart[2], Frédérique Gancel[1], Marlène Chollet-Imbert[1] and Philippe Jacques[1*]

*Laboratory of Microbial Bioprocesses[1], Polytech'Lille, University of Sciences and Technologies of Lille, F-59655 Villeneuve d'Ascq Cedex, France, Centre Wallon de Biologie Industrielle[2], University of Liege, B40, B-4000 Liège, Belgium, and Unité de Chimie Biologique Industrielle[3], Agricultural University of Gembloux, B-5030 Gembloux, Belgium*

[*]Corresponding author. Mailing address: Laboratory of Microbial Bioprocesses (LABEM), Polytech'Lille, University of Sciences and Technologies of Lille, Avenue du Professeur Langevin, F-59655 Villeneuve d'Ascq Cedex, France Phone: + 33 3 28 76 74 40 Fax: + 33 3 28 76 74 01. E-mail : philippe.jacques@polytech-lille.fr

**Abstract**

A *Bacillus subtilis* derivative was obtained from strain ATCC 6633 by the replacement of the native promoter of the mycosubtilin operon by a constitutive one, originating from the replication gene *repU* of the *Staphylococcus aureus* plasmid pUB110. The recombinant strain named BBG100 produced up to 15-fold more mycosubtilin than the wild-type. This overproducing phenotype was related to the enhancement of antagonistic activities against several yeasts and pathogenic fungi. Hemolytic activities were also clearly increased in the modified strain. Mass spectrometry analyses of enriched mycosubtilin extracts showed similar patterns of lipopeptides for BBG100 and the wild-type. Interestingly, these analyses also revealed a new form of mycosubtilin which is more easily detected in BBG100 sample. When tested for its biocontrol potential, the wild-type strain ATCC 6633 was almost ineffective at reducing *Pythium* infection of tomato seedlings. However, treatment of seeds with the BBG100 overproducing strain resulted in a marked increase of the germination rate of plantlets. This protective effect afforded by mycosubtilin overproduction was also visualized by the significantly higher fresh weight of emerging seedlings treated with BBG100 compared to controls or those inoculated with the wild-type strain.

**INTRODUCTION**

Members of the *Bacillus subtilis* family produce a wide variety of antibacterial and antifungal antibiotics. Some of them like subtilin (41), subtilosin A (2), TasA (34) and sublancin (27) are of ribosomal origin but others such as bacilysin, chlorotetain, mycobacillin (41), rhizocticins (19), bacillaene (28), difficidin (40) and lipopeptides from the surfactin, iturin and fengycin families (41) are formed by non-ribosomal peptide synthetases and/or polyketide synthases. The later are amphiphilic cyclic peptides composed of seven (surfactins and iturins) or ten α-amino acids (fengycins) linked to one unique β-amino (iturins) or β-hydroxy (surfactins and fengycins) fatty acid. The length of this fatty acid chain may vary from $C_{13}$ to $C_{16}$ for surfactins, from $C_{14}$ to $C_{17}$ for iturins and from $C_{14}$ to $C_{18}$ in the case of fengycins. Different homologous compounds for each lipopeptide family are thus usually co-produced (1, 16). Iturins and fengycins display a strong antifungal activity and are inhibitory for the growth of a wide range of plant pathogens (11, 17, 20, 22, 35). Surfactins are not fungitoxic by themselves but retain some synergistic effect on the antifungal activity of iturin A (23).

*B. subtilis* ATCC 6633 produces subtilin (21), subtilosin (33), rhizocticin (19) and two lipopeptides: surfactin and mycosubtilin, a member of the iturin family (21). Production of surfactin requires the *srf* operon encoding the three subunits of surfactin synthetase that catalyse the thiotemplate mechanism of non-ribosomal peptide synthesis to incorporate the seven amino acids into the surfactin lipopeptide. The mycosubtilin gene cluster consists of four ORFs designated *fenF*, *mycA*, *mycB* and *mycC* controlled by the same promoter $P_{myc}$ (Fig. 1) (9). The subunits encoded by the three *myc* genes contain the seven modules necessary to synthesize the peptidic moiety of mycosubtilin. The N-terminal multifunctional part of *mycA* shows strong homology with fatty acid and polyketide synthases.

The production of surfactin is activated by a regulatory system coupled to the accumulation of cell-derived extracellular signals at the end of the exponential growth (7) while iturin synthesis is induced during the stationary phase (16).

Among the biological control alternatives to chemical pesticides used for reducing plant diseases, the application of non-pathogenic soil bacteria living in association with plant roots is promising. Treatment with these beneficial organisms was in many cases associated with reduced plant diseases in greenhouse and field experiments. These bacteria can antagonize fungal pathogens by competing for niche and nutriments, by producing low molecular weight fungitoxic compounds and extracellular lytic enzymes and more indirectly by stimulating the defensive capacities of the host plant (10, 26, 30, 35). On the basis of the wide diversity of powerful antifungal metabolites that can be synthesized by *B. subtilis*, it was suggested that antibiotic production by these strains played a major role in plant disease suppression (4, 32, 35, 38). These bacteria were reported to be effective at controlling many plant or fruit diseases caused either by soilborne, aerial or post-harvest pathogens (4, 22, 35, 37, 39). Some of these strains are currently used in commercially available biocontrol products (3, 5). However most of the studies have primarily focused on the degree of disease reduction and mechanisms of suppression in soil have not been as extensively investigated.

In this study, the native promoter of the mycosubtilin operon from *B. subtilis* ATCC 6633 was replaced by the promoter $P_{repU}$ from the staphylococcal plasmid pUB110 which was proved to be strong and constitutive in *Bacillus subtilis* (36). Growth and lipopeptide production by the derivative were compared to the wild-type as well as their antimicrobial and hemolytic activities. The effect of the early overproduction of mycosubtilin in the biocontrol of damping-off caused by *Pythium aphanidermatum* on tomato seedlings was also evaluated.

## MATERIALS AND METHODS

**Bacterial strains, plasmids, and growth conditions.** The microorganisms and plasmids used in this study are listed in Table 1. *B. subtilis* strains were grown at 30°C in either Landy medium (20) or medium 863 (1). *E. coli* DH5α was cultured at 37°C in Luria-Bertani medium (LB) supplemented, when required, with various antibiotics: ampicillin (Ap; Sigma, St. Louis, MO; 50 µg ml$^{-1}$), neomycin (Nm; Serva, Heidelberg, Germany; 20 µg ml$^{-1}$) and streptomycin (Sm; Sigma; 25 µg ml$^{-1}$). The yeast strains were grown at 28°C in medium 863 (1) and the fungal strains were cultured at 30°C on potato-dextrose-agar (PDA; Biokar Diagnostics, Beauvais, France) .

**Molecular biology procedures.** Total genomic DNA was extracted from *B. subtilis* ATCC 6633 and purified using the genomic tips 20/G together with the corresponding buffers purchased from Qiagen (Hilden, Germany). Plasmid DNAs were prepared from *E. coli* using either the Miniprep Spin or Maxiprep kits (Qiagen). Screening for hybrid plamids within various *E.coli* transformants was done by the "boiling" procedure of Holmes and Quigley (13). Restriction endonucleases digestions, ligation and transformation of *E. coli* by the CaCl$_2$ thermal shock followed standard procedures (31). *B. subtilis* ATCC 6633 was transformed by electroporation according to the method of Dennis and Sokol (8).

For the construction of the pUC19-derived plasmid dedicated to promoter exchange by homologous recombination in *B. subtilis*, the *pbp* and *fenF* fragments were generated by PCR using Taq polymerase "Arrow" from Qbiogene (Montreal, Canada). The primers were designed according to the published sequence of the mycosubtilin operon from strain ATCC 6633 (PubMed Nucleotide AF184956) (9). The primers were: (i) for *pbp*, the forward one, 5'-TTAGAAGA<u>GCATGC</u>AAAAATG-3' (the underlined artificial *Sph*I site was generated by substitution of the two bases in bold characters); and the reverse one, 5'-CCCTCCAATCTTTTCGAACG-3'; and (ii) for *fenF*, the forward one, 5'-GACATGTATCCGT<u>TC**T**AG**A**</u>AGATTG-3' (the underlined artificial *Xba*I site was generated by substitution of the two bases in bold characters); and the reverse one, 5'-ATCGGCCATTCAGCATCTC-3'). PCR conditions consisted of an initial denaturation step at 95°C for 2 min, followed by 30 cycles of (i) 30 s at 95°C; (ii) 30 s at 45°C; and (iii) 30 s at 70°C. The final extension step was at 70°C for 2 min.

The two PCR-generated cassettes were purified from 2% agarose gels using the QIAquick kit (Qiagen), treated with proteinase K (50 µg ml$^{-1}$) for 1 h at 37°C and subjected to deproteinisation using a phenol/chloroform procedure. The *fenF* fragment was *Xba*I and *Bsp*E1 double digested and introduced between the *Xba*I and *Xma*I sites of pUC19 to yield pBG101. After *Sph*I and *Mph*1103I double digestion, the *pbp* fragment was inserted within *Sph*I and *Pst*I sites of pUC19 generating pBG102. Then, after *Eco*RI and *Sal*I double digestion, the *fenF* fragment was inserted at the corresponding sites of pBG102. The resulting construct was named pBG103. After *Xba*I digestion, the P$_{repU}$ –*neo* fragment was extracted from pBEST501 (15) and inserted into the *Xba*I site of pBG103. This construct, named pBG106 (Fig. 1), was then used to transform *B. subtilis* ATCC 6633, which was plated on LB agar containing neomycin to select recombinants and incubated at 37°C.

**Lipopeptide purification and identification.** Cultures were centrifuged at 15,000 x g for 1 h at 4°C. For lipopeptide extraction, 1-ml samples of supernatants were purified on C$_{18}$ Maxi-Clean cartridges (Alltech, Deerfield, IL) according to the recommendations of the supplier. Lipopeptides were eluted with 5 ml of pure methanol (HPLC grade, Acros Organics, Geel, Belgium). The extract was brought to dryness and the residue dissolved in methanol (200 µl) before analysis by high-performance liquid chromatography (HPLC) using a C$_{18}$ column (5 μm, 250 x 4.6 mm, VYDAC 218 TP, Hesperia, CA). Each family of lipopeptides was separately analyzed with the solvent system acetonitrile/water/trifluoroacetic acid (TFA) which was used in the proportions (40:60:0.5; vol/vol/vol) and (80:20:0.5, vol/vol/vol) for iturins and surfactins respectively. 20-μl samples were injected and compounds were eluted at a flow rate of 1ml min$^{-1}$. Purified iturins and surfactins were purchased from Sigma (Saint Louis, MO). Retention time and second derivatives of UV-visible spectra (Diode Array Waters PDA 996, Millenium Software) of each peak were used to identify the eluted molecules.

Lipopeptide extracts were further analyzed by MALDI-TOF MS. A saturated solution of CHCA (α-cyano-4-hydroxy-cinnamic acid) was prepared in a 3:1 (vol/vol) solution of $CH_3CN/H_2O$ 0.1% TFA. The cell culture supernatant was diluted tenfold with CHCA-saturated solution. 0.5 μl of this solution was deposited on the target. Measurement was performed using the UV laser desorption time-of-flight mass spectrometer Bruker Ultraflex tof (Bruker Daltonics), equipped with a pulsed nitrogen laser ($\lambda = 337$ nm). The analyzer was used at an acceleration voltage of 20 kV. Samples were measured in the reflectron mode.

**Evaluation of antimicrobial and hemolytic activities.** Supernatants from *B. subtilis* cultures obtained from various media were filter-sterilized through 0.2 μm pore size membranes and treated or not for 1 h at 37°C with protease (Sigma, type XIV; at 10 μg ml$^{-1}$, final concentration) to neutralize subtilin and subtilosin activities.

Antimicrobial activities of supernatant samples from both wild-type and modified strains were tested on plate bioassays. Bacterial and yeast strains to be tested were grown in LB or 863 medium, respectively. Overnight bacterial cultures (2 ml) were diluted (10$^{-2}$) and inoculated by flooding a 2-ml volume on LB plates. The excess of liquid was withdrawn and the plates were allowed to dry under a laminar flow hood for 30 min. In tests performed with yeast strains, 4 ml of semi-solidified 863 medium (0.8% agar) containing 100 µl of diluted cell suspension ($10^{-1}$) were spread onto 863 plates. In both cases, 200 µl of supernatant samples were deposited in 10 mm diameter wells created in the solidified media using sterile glass tubes. The plates were incubated at either 30°C or 37°C depending on the strain to be tested. A similar method was used to test supernatant samples for their antifungal activities against filamentous fungi. Mycelial plugs (5 mm) were deposited in the center of the plates, at equal distances from the wells. Plates were incubated at 28°C and inhibition zones were measured after 1 to 3 days. To evaluate hemolytic activity of the various supernatants, 200 µl-samples were dispensed in wells made in blood agar plates (with 5% defibrinated sheep blood; Eurobio, Les Ulis, France). Hemolytic activity was visualized by the development of a clear halo around the wells after incubation at 37°C. In all cases, two replicate plates were used for each strain on each medium and the experiment was repeated once.

**Determination of MIC.** Serial half-dilutions of filter-sterilized culture supernatants, containing known concentrations of mycosubtilin, were performed up to 1/1024 using 863 medium. After inoculation with 100 µl of diluted *S. cerevisiae* culture (about $10^5$ cell ml$^{-1}$), the test tubes were incubated at 30°C. The MIC was determined by taking into account the higher dilution where no growth of the test organism was visible.

**Biocontrol assays with tomato.** For the preparation of bacterial inoculum, the *Bacillus* strains were grown at 30°C for 24 h in Landy medium. Cells were harvested by centrifugation at 35,000 x *g* for 20 min and the cell pellet was washed twice with sterile saline water (0.85% NaCl). Vegetative cell suspensions were then diluted in order to obtain the desired bacterial concentration for seed treatment. The origin of the fungal pathogen *Pythium* *aphanidermatum*, its maintenance and the preparation of suspensions used in the bioassays were previously described (24).

In the damping-off assays, tomato seeds (*Lycopersicon esculentum* L. cv Merveille des Marchés) were germinated in a peat substrate (Brill Substrate GmbH & Co KG, Georgsdorf, Germany) hereafter referred to as "soil". Prior to sowing, seeds were washed three times (for 5 min each) with sterile distilled water and soaked for 10 min in the appropriate bacterial suspension at a concentration of approximately $4 \times 10^8$ CFU ml$^{-1}$ or in NaCl 0.85% in the case of control plants. In every experiment, 200 seeds were used for each treatment. They were sown in large plastic trays containing soil previously infected with *P. aphanidermatum* by mixing with a suspension of mycelial fragments. Final concentration of the pathogen in the substrate for plant growth was $10^5$ propagules g$^{-1}$ of soil dry weight. The trays were incubated in a growth cabinet set to maintain the temperature at 28°C, a 95%-relative humidity and a photoperiod of 16 h. Seedling emergence was recorded after 12 days and the number of healthy plantlets was reported to the number of seeds.

**RESULTS**

**Construction of the BBG100 mutant by allelic exchange.** Several transformation experiments of *B. subtilis* ATCC6633 with the pBG106 led to the isolation of 15 Nm$^R$ colonies. Genomic DNA of these clones and the wild-type strain was purified. Direct observation of the restriction endonucleases (HindIII and PstI) profiles did not point out any major difference (data not shown). The replacement of the natural promoter by the constitutive one P$_{repU}$ associated with the *neo* gene was demonstrated by PCR amplification of genomic DNA with the εpbp forward and the εfenF reverse primers. For one of the different tested colonies, a ~ 2.8 kb fragment was obtained instead of the ~1.5 kb fragment obtained with the wild-type. The corresponding modified strains, named BBG100, was further compared to the wild-type for its lipopeptide production level and biological activities.

**Mycosubtilin overproduction by BBG100**. Mycosubtilin production was followed up upon growth of both strains in agitated Erlenmeyer flask and 3-L bioreactor during 3 days (Table 2). Although the absolute level of mycosubtilin is different in shake flasks versus the bioreactor, 12- to 15-fold increases were observed after 72 h in BBG100 culture supernatant in the two different growth conditions. The higher productivity of mycosubtilin was obtained in flask with BBG100 (63.6 mg/g of cells). As expected, surfactin synthesis was not affected by the promoter replacement since production levels in BBG100 and wild-type remained similar under both growth conditions. The lower concentrations found in bioreactors compared to those obtained in shake flasks are probably due to the low aeration rate used in the bioreactor in order to limit liquid extraction by foaming. This resulted in lower oxygen transfer compared to well-agitated flasks and thereby in reduced lipopeptide production rate since the synthesis of these molecules is positively influenced by oxygen (16).

Time course evolution of biomass concentration and pH value during the 72-h growth in fermentor were also closely similar for both strains. Typically, acidification of the medium was observed during the early exponential growth phase being due to the production of organic acids from glucose. This was followed by a neutralization step during the second growth phase related to the consumption of these acids and by a slight alcalinization step due to the use of glutamic acid as carbon source by the cells (data not shown). It is thus likely that BBG100 conserved a physiological behavior similar to the wild-type.

Analysis of lipopeptide production during the first 8 h of growth in bioreactor revealed a early synthesis of mycosubtilin by BBG100 (Fig 2). Significant amounts of mycosubtilin were already produced after 4 h of incubation when the cells entered the exponential growth phase. Despite similar biomass level, mycosubtilin production by the wild–type was not observed over the first 8 h as expected since the synthesis of these compounds is known to occur only at the beginning of the stationary phase.

MALDI-TOF mass spectrometry analyses of lipopeptide extracts allowed the identification of several homologues for surfactins and mycosubtilins produced by both strains (Fig. 3). Signal attributions to protonated form of mycosubtilin and surfactin and their $Na^+$ and $K^+$ adducts are summarized in Table 3. However, MS peaks showing a higher intensity were detected in the extract from the BBG100: a signal at m/z 1095.65 which corresponds to the $M+K^+$ ion of the C-15 homologue of mycosubtilin and, more interestingly, a signal at m/z 1137.7 which cannot be attributed to known ions of surfactin or mycosubtilin.

**Biological activities.** BBG100 and wild-type were compared for their antagonistic properties against a wide range of microorganisms. Supernatants from both strains did not inhibit the growth of *E. chrysanthemi*, *E. coli* and *P. aeruginosa* even upon tenfold concentrations. When tested on *M. luteus*, however, the two supernatants generated similar growth inhibition zones that completely disappeared upon treatment with protease type XIV which neutralize bacteriocin-like activities. By contrast, BBG100 culture supernatant induced growth inhibition zones significantly greater than those observed for the wild-type one when tested against three phytopathogenic fungi, *B. cinerea*, *F. oxysporum* and *P. aphanidermatum* and two yeasts, *P. pastoris* and *S. cerevisiae* (Table 4). Protease treatment of the supernatants slightly reduced the antifungal activity against *P. aphanidermatum*.

Serial dilutions of culture supernatant from both strains were tested independently for their inhibitory effect toward growth of *S. cerevisiae*. Eight-fold higher dilution of BBG100 supernatant was necessary to obtain the minimal inhibitory concentration (MIC) of mycosubtilin as compared to wild-type. In both cases, this MIC was determined as 8 µg ml$^{-1}$. It confirmed that antagonistic activity against yeast of both supernatants was essentially due to mycosubtilin.

When tested for its lytic activity on blood corpuscles, the supernatant from BBG100 yielded greater hemolytic areas than those observed for the wild-type (Fig. 4).

**Protection against *Pythium* damping-off of tomato seedlings.** Biocontrol assays were conducted in the tomato/*Pythium* pathosystem to compare the ability of the wild-type strain ATCC 6633 with that of BBG100 at reducing seedling infection. As shown in Table 5, pre-treatment of tomato seeds with vegetative cells of the wild-type strain failed to provide any protective effect but appeared to be conducive to disease development. However inoculation with the lipopeptide-overproducing derivative prior to planting led to enhanced seedling emergence that was consistently observed over four independent experiments while strong differences were observed in disease incidence. Whatever they were previously inoculated with the wild-type or with the BBG100 strain or none (healthy control), the germination rate of seeds in the absence of pathogen did not vary significantly and was in most of the cases comprised between 90% and 95% (Table 5). The protective effect of BBG100 was also illustrated by an increase in the size and vigor of emerging plantlets compared to disease controls or plants inoculated with the wild-type (Fig. 5). In one representative experiment, the mean value for fresh weight of individual plants (aerial part, harvested after 18 days of incubation) was significantly higher following seed treatment with the BBG100 strain (0.79 g/plant) than for non-bacterized plants (0.31 g/plant) or for those inoculated with the wild-type strain 6633 (0.23 g/plant).

**DISCUSSION**

In this work, we replaced the native promoter of the mycosubtilin operon of *B. subtilis* ATCC 6633 by a constitutive one which governs the replication gene *repU* from the *S. aureus* plasmid pUB110. This led to the isolation of the BBG100 derivative displaying a 15-fold increase in mycosubtilin production rate. This $P_{repU}$ promoter was previously reported to enhance the biosynthesis of iturin A, another antifungal lipopeptide structurally very close to mycosubtilin, by about three times in *B. subtilis* RB14 (36).

When tested against different bacteria, yeast and fungi, the supernatant of wild-type strain only showed a very good antagonistic activity against *M. luteus*. This activity which was also detected with the supernatant of the modified strain, completely disappeared upon pre-treatment with protease. This antibiotic activity could thus be attributed to some protease sensitive compounds like subtilin and subtilosin, known to be produced by this strain (21, 33). The very weak antifungal activity displayed by the wild-type strain suggested that rhizocticins and mycosubtilin are produced in very low amounts. The slight reduction of antagonistic activity against *P. aphanidermatum* observed after proteolytic treatment could result from amino acids or oligopeptides liberated by the treatment and known to neutralize biological activity of rhizocticin (19). By contrast, $P_{repU}$–governed mycosubtilin overproduction in *B. subtilis* BBG100 led to clearly enhanced fungitoxic activities showing that this lipopeptide plays a crucial role in the antagonism developed by the strain.

When applied to seed or mixed with soil, some *B. subtilis* strains were reported to provide crop protection mostly by direct control of soilborne pathogens through an efficient production of various fungitoxic metabolites (3, 29, 32). By using the tomato/*P. aphanidermatum* pathosystem, this study demonstrates that overproduction of mycosubtilin by *B. subtilis* ATCC 6633 may confer some biocontrol potential to a strain naturally not active at protecting plants. Considering the mean values calculated from pooled data, the germination rate of plants treated with the mycosubtilin overproducer increased by 31% when compared to control seeds and by 48% as compared to the wild-type. As mycosubtilin displays a strong antifungal activity *in vitro* against *P. aphanidermatum*, it is obvious that the 15-fold higher *in vitro* production rate of this compound is tightly involved in the protective effect developed *in vivo* by the modified strain. An early and higher production of the lipopeptides probably enhances the biological effect of the strain by immedially reducing plant pathogen growth. The role played by these molecules is reinforced by the fact that other possible biocontrol mechanisms are seemingly not concerned. For example, some *B. subtilis* strains were reported to reduce disease incidence indirectly by triggering systemic resistance in the plant (25). We have performed some experiments with tomatoes pre-inoculated at the root level with either the wild-type ATCC 6633 or the mycosubtilin overproducer derivative before challenge with the pathogen *B. cinerea* on leaves. Such procedure is used to reveal disease suppression due to induction of resistance in the host plant by the bacteria. However, none of the strains was able to develop some protective effect under these conditions showing that they do not retain any plant resistance inducing activity (data not shown). In the same line, growth-promotion activity *sensu stricto* could also probably not be evoked to explain the beneficial effect of the mycosubtilin overproducer. Size and robustness of plants inoculated with the modified strain were higher than those of disease controls and very similar to untreated controls when grown in a soil not infested with the pathogen (data not shown). By contrast with its overproducing derivative, the wild type strain ATCC 6633 did not display any protective effect on tomato seedlings. Surprisingly the strain 6633 even appeared to be conducive to the disease. However, when grown in the absence of pathogen, tomato plantlets inoculated with the wild-type were similar to the control plants suggesting that the strain did not develop any phytotoxic effect *per se*.

Mass spectrometry analyses of supernatants from *B. subtilis* ATCC 6633 and BBG100 revealed the presence of two main molecular ions corresponding to the homologous mycosubtilins with $C_{16}$ or $C_{17}$ fatty acid chains. These homologues are considered as more biologically active compared to the iturins that bear shorter hydrocarbon side chain ($C_{14}$-$C_{15}$) (11). It was shown that fungitoxicity increases with the number of carbon in the fatty acid chain i.e. $C_{17}$ homologues are 20-fold more active than the $C_{14}$ forms. This is also evidenced by the similarity of *in vitro* antagonistic activity developed by BBG100 and the one shown by other *Bacillus* strains producing higher amounts of iturinic compounds but with shorter fatty acid chains (16, 35).

The overproduction of mycosubtilin in the BBG100 derivative is also accompanied by qualitative changes in the pattern of lipopeptides. Interestingly a signal at m/z 1137.7 was clearly enhanced. The corresponding compound is probably structurally similar to iturins since it followed the purification of mycosubtilin. In addition, it should correspond to a $K^+$ adduct since MS/MS analysis did not yield any fragmentation (data not shown). Bacillomycin F with a $C_{17}$ fatty acid chain is the sole iturin form that could correspond to this molecular weight. However, a single insertion of the new promoter was confirmed in the mycosubtilin operon. So, overexpression of bacillomycin synthetases is obviously not involved. This signal could thus preferably be attributed to a modified mycosubtilin with either a $C_{18}$ chain of fatty acid or a peptide moiety containing a Thr instead of a Ser. In both cases, this molecule represents a new form of mycosubtilin. Indeed, such long fatty acid chain was never encountered in iturin-like lipopeptides and amino acid residue replacement has never been demonstrated with iturin derivatives. However this last phenomenon may occur as shown in the case of the non-ribosomal surfactin synthetase which possesses adenylation domains able to activate different amino acid residues with similar side chains (18). Similarly, the mycobactin synthetase contains an adenylation domain that may recognize both L-serine and L-threonine (6). Such a low specificity could thus also be observed in mycosubtilin synthetase. Further structural investigations are being performed to confirm this hypothesis.

**ACKNOWLEDGEMENTS**

This work received financial support from the Université des Sciences et Technologies de Lille, the Région Nord-Pas de Calais, the Fonds Européen pour le Développement de la Recherche and the National Funds for Scientific Research (F.N.R.S., Belgium, Program F.R.F.C. n° 2.4.570.00).

# REFERENCES

1. **Akpa E., P. Jacques, B. Wathelet, M. Paquot, R. Fuchs, H. Budzikiewicz, and P. Thonart**. 2001. Influence of culture conditions on lipopeptide production by *Bacillus subtilis*. Appl. Biochem. Biotechnol. **91-93:**551-561.

2. **Babasaki, K., T. Takao, Y. Shimonishi, and K. Kurahashi.** 1985. Subtilosin A, a new antibiotic peptide produced by *Bacillus subtilis* 168: isolation, structural analysis, and biogenesis. J. Biochem. (Tokyo) **98:**585-603.

3. **Backman, P. A., M. Wilson, and J. F. Murphy.** 1997. Bacteria for biological control of plant diseases. Lewis Publishers, Boca Raton, Florida.

4. **Bais, P. B., R. Fall, and J. M. Vivanco.** 2004. Biocontrol of *Bacillus subtilis* against infection of *Arabidopsis* roots by *Pseudomonas syringae* is facilitated by biofilm formation and surfactin production. Plant Physiol. **134:**307-319.

5. **Brannen, P. M., and D. S. Kenney.** 1997. Kodiak$^R$-a successful biological-control product for suppression of soil-borne plant pathogens of cotton. J. Ind. Microbiol. Biotechnol. **19:**169-171.

6. **Challis, G. L., J. Ravel, and C. A. Townsend.** 2000. Predictive, structure-based model of amino-acid recognition by nonribosomal peptide synthetase adenylation domains. Chem. Biol. **7:**211-224.

7. **Cosby, W. M., D. Vollenbroich, O. H. Lee, and P. Zuber.** 1998. Altered *srf* expression in *Bacillus subtilis* resulting from changes in culture pH is dependent on the SpoOK oligopeptide permease and the ComQX system of extracellular compounds. J. Bacteriol. **180:**1438-1445.

8. **Dennis, J. J., and P. A. Sokol**. 1995. Electrotransformation of *Pseudomonas*. p. 125-133. *In* Nickoloff, J. A. (ed.), Methods in Molecular Biology. Electroporation Protocols for Microorganisms, Vol. 47,. Humana Press Inc, Totowa, NJ.

9. **Duitman, E. H., L. W. Hamoen, M. Rembold, G. Venema, H. Seitz, W. Saenger, F. Bernhardt, M. Schmidt, C. Ulrich, T. Stein, F. Leenders, and J. Vater**. 1999. The mycosubtilin synthetase of *Bacillus subtilis* ATCC 6633: a multifunctional hybrid between a peptide synthetase, an amino transferase, and a fatty acid synthase. Proc. Natl. Acad. Sci. USA **96**:13294-13299.

10. **Handelsman, J., and E. V. Stabb.** 1996. Biocontrol of soilborne plant pathogens. Plant Cell **8:**1855-1869.

11. **Hbid, C.** 1996. Ph.D. thesis. University of Liege, Belgium.

12. **Höfte, M., S. Buysens, N. Koedam, and P. Cornelis.** 1993. Zinc affects siderophore-mediated high affinity iron uptake systems in the rhizosphere *Pseudomonas aeruginosa* 7NSK2. Biometals **6:**85-91.

13. **Holmes, D. S., and M. Quigley.** 1981. A rapid boiling method for the preparation of bacterial plasmids. Anal. Biochem. **114:**193-197.

14. **Hugouvieux-Cotte-Patat N., H. Dominguez, and J. Robert-Baudouy.** 1992. Environmental conditions affect transcription of the pectinase genes of *Erwinia chrysanthemi*. J. Bacteriol. **174:**7807-7818.

15. **Itaya, M., K. Kondo, and T. Tanaka**. 1989. A neomycin resistance gene cassette selectable in a single copy state in the *Bacillus subtilis* chromosome. Nucl. Acids Res. **17:**4410.

16. **Jacques, P., C. Hbid, J. Destain, H. Razafindralambo, M. Paquot, E. De Pauw, and P. Thonart.** 1999. Optimization of biosurfactant lipopeptide production from *Bacillus subtilis* S499 by Plackett-Burman design. Appl. Biochem. Biotechnol. **77:**223-233.

17. **Koutmousi, A., X. H. Chen, A. Henne, H. Liesegang, G. Hitzeroth, P. Franke, J. Vater, and R. Borriss.** 2004. Structural and functional characterization of gene clusters directing nonribosomal synthesis of bioactive cyclic lipopeptides in *Bacillus amyloliquefaciens* strain FZB42. J. Bacteriol. **186:**1084-1096.

18. **Kowall, M., J. Vater, B. Kluge, T. Stein, P. Franke, and D. Ziessow.** 1998. Separation and characterization of surfactin isoforms produced by *Bacillus subtilis* OKB 105. J. Colloid Interf. Sci. **204:**1-8.

19. **Kugler, M., W. Loeffler, C. Rapp, A. Kern, and G. Jung.** 1990. Rhizocticin A, an antifungal phosphono-oligopeptide of *Bacillus subtilis* ATCC 6633: biological properties. Arch. Microbiol. **153:**276-281.

20. **Landy, M., G. H. Warren, S. B. Roseman, and L. G. Golio.** 1948. Bacillomycin, an antibiotic from *Bacillus subtilis* active against pathogenic fungi. Proc. Soc. Exp. Biol. Med. **67:**539-541.

21. **Leenders, F., T. H. Stein, B. Kablitz, P. Franke, and J. Vater.** 1999. Rapid typing of *Bacillus subtilis* strains by their secondary metabolites using matrix assisted laser-desorption/ionization mass spectrometry of intact cells. Rapid Commun. Mass Spectrom. **13:**943-949.

22. **Leifert, C., H. Li, S. Chidburee, S. Hampson, S. Workman, D. Sigee, H. A. Epton, and A. Harbour.** 1995. Antibiotic production and biocontrol activity by *Bacillus subtilis* CL27 and *Bacillus pumilus* CL45. J. Appl. Bacteriol. **78:**97-108.

23. **Maget-Dana R., L. Thimon, F. Peypoux, and M. Ptak.** 1992. Surfactin/iturin A interactions may explain the synergistic effect of surfactin on the biological properties of iturin A. Biochimie **74:**1047-1051.

24. **Ongena, M., F. Daayf, P. Jacques, P. Thonart, N. Benhamou, T. C. Paulitz, and R. R. Belanger.** 2000. Systemic induction of phytoalexins in cucumber in response to treatments with fluorescent pseudomonads. Plant Pathol. **49:**523-530.

25. **Ongena, M., F. Duby, E. Jourdan, J. Dommes, and P. Thonart.** Appl Microbiol Biotechnol., in press.

26. **Ongena, M., F. Duby, F. Rossignol, M. L. Fauconnier, J. Dommes, and P. Thonart.** 2004. Stimulation of the lipoxygenase pathway is associated with systemic resistance induced in bean by a non-pathogenic *Pseudomonas* strain. Mol. Plant-Microbe Interact. **17:**1009-1018.

27. **Paik, S. H., A. Chakicherla, and J. N. Hansen.** 1998. Identification and characterization of the structural and transporter genes for, and the chemical and biological properties of sublancin 168, a novel lantibiotic produced by *Bacillus subtilis* 168. J. Biol. Chem. **273:**23134-23142.

28. **Patel P.S., S. Huang, S. Fisher, D. Pirnik, C. Aklonis, L. Dean, E. Meyers, P. Fernandes, and F. Mayerl**. 1995. Bacillaene, a novel inhibitor of procaryotic protein synthesis produced by *Bacillus subtilis*: production, taxonomy, isolation, physico-chemical characterization and biological activity. J. Antibiot. (Tokyo) **48:**997-1003.

29. **Phae, C. G., M. Shoda, and H. Kubota.** 1990. Suppressive effect of *Bacillus subtilis* and its products on phytopathogenic microorganisms. J. Ferm. Bioeng. **69:**1-7.

30. **Raupach, G. S., and J. W. Kloepper.** 1998. Mixtures of plant growth-promoting rhizobacteria enhance biological control of multiple cucumber pathogens. Phytopathology **88:**1158-1164.

31. **Sambrook, J., and D.W. Russel.** 2001. Molecular cloning: A laboratory manual, 3rd edn. New York: Cold Spring Harbor Laboratory.

32. **Shoda, M.** 2000. Bacterial control of plant diseases. J. Biosci. Bioeng. **89:**515-521.

33. **Stein, T., S. Düsterhus, A. Stroh, and K.D. Entian.** 2004. Subtilosin production by two *Bacillus subtilis* subspecies and variance of the *sbo-alb* cluster. Appl. Environ. Microbiol. **70:**2349-2353.

34. **Stover, A. G. and A. Driks.** 1999. Secretion, localization, and antibacterial activity of TasA, a *Bacillus subtilis* spore-associated protein. J. Bacteriol. **181:**1664-1672.

35. **Touré, Y., M. Ongena, P. Jacques, A. Guiro, and P. Thonart.** 2004. Role of lipopeptides produced by *Bacillus subtilis* GA1 in the reduction of grey mould disease caused by *Botrytis cinerea* on apple. J. Appl. Microbiol. **96:**1151-1160.

36. **Tsuge, K., T. Akiyama, and M. Shoda**. 2001. Cloning, sequencing, and characterization of the iturin A operon. J. Bacteriol. **183**:6265-6273.

37. **Whipps, J. M.** 2001. Microbial interactions and biocontrol in the rhizosphere. J. Exp. Bot. **52:**487-511.

38. **Yoshida, S., S. Hiradate, T. Tsukamoto, K. Hatakeda, and A. Shirata.** 2001. Antimicrobial activity of culture filtrate of *Bacillus amyloliquefaciens* RC-2 isolated from mulberry leaves. Phytopathology **91:**181-187.

39. **Yu, G. Y., J. B. Sinclair, G. L. Hartman, and B. L. Bertagnolli.** 2002. Production of iturin A by *Bacillus amyloliquefaciens* suppressing *Rhizoctonia solani*. Soil Biol. Biochem. **34:**955-963.

40. **Zimmerman S. B., C. D. Schwartz, R. L. Monaghan, B. A. Pelak, B. Weissberger, E. C. Gilfillan, S. Mochales, S. Hernandez, S. A. Currie, E. Tejera,** *et al.* 1987. Difficidin and oxydifficidin: novel broad spectrum antibacterial antibiotics produced by *Bacillus subtilis*. I. Production, taxonomy and antibacterial activity. J. Antibiot. (Tokyo) **40:**1677-1681.

41. **Zuber P., M. M. Nakano, and M. A. Marahiel.** 1993. Peptide antibiotics, p. 897-916. *In* A. L. Sonenshein, J. A. Hoch, and R. Losick (ed.), *Bacillus subtilis* and other gram-positive bacteria. American Society for Microbiology, Washington, D.C.

**FIGURE LEGENDS**

FIG. 1. Replacement in *B. subtilis* ATCC 6633 of the original P*myc* promoter by the P$_{repU}$-*neo* cassette using homologous recombination between genomic DNA of the strain and the hybrid plasmid pBG106. (A) Recognition of homologous regions located (i) after the termination region of the *pbp* gene (coding for a penicillin-binding protein) situated upstream the mycosubtilin operon (for convenience, the cassette generated by PCR in this region was named ε"*pbp*"); and (ii) immediately downstream the p*myc* promoter (cassette ε*fenF*). The four genes *fenF*, *mycA*, *mycB* and *mycC* constitute the mycosubtilin operon, and code for a malonyl-CoA transacylase and three peptide synthetases, respectively; *yngL*, gene coding for an unknown function; P*repU*, promoter of the replication gene of pUB110; and *neo*, gene conferring resistance to neomycin/kanamycin from pUB110 (15); (∗) site newly created after ligation between the *Bsp*EI and *Xma*I compatible cohesive ends. (B) Construct obtained within the genomic DNA of the strain following homologous recombination (generated by the inability of pUC19 to replicate in *Bacillus* spp., together with the selective pressure for resistance to neomycin); the mycosubtilin operon became under control of the P$_{repU}$ constitutive promoter.

FIG. 2. Early stage of growth (solid symbols) and mycosubtilin (open symbols) production of *B. subtilis* ATCC 6633 (□) and its BBG100 derivative (Δ) in bioreactor.

FIG. 3. MALDI-TOF spectra of lipopeptides produced by *B. subtilis* ATCC 6633 (A) and BBG100 (B).

FIG. 4. Hemolytic activities of supernatants obtained after growth of the wild-type strain in Landy (a) or 863 (c) medium; and of the strain BBG100 in Landy (b) or 863 (d) medium.

FIG. 5. Illustration of plantlets obtained from seeds treated either with the wild-type *B. subtilis* strain ATCC 6633 (A), with its mycosubtilin overproducing derivative BBG100 (B) or with water (C) (disease control) in *P. aphanidermatum*-infested soil, 18 days after sowing.

TABLE 1. Strains and plasmids

| Strain or plasmid | Description[a] | Source or reference |
|---|---|---|
| Bacterial strains | | |
| *Escherichia coli* DH5α | Φ80d*lacZ*ΔM15 *rec*A1 *end*A1 *gyr*A96 *thi*-1 *hsd*R17 ($r_k$-, $m_k$+) *sup*E44 *rel*A1 *deo*R, Δ(*lac*ZYA-*arg*F)U169 *pho*A | Promega, Madison, WI |
| *Bacillus subtilis* ATCC 6633 | Produces mycosubtilin, surfactin, subtilin, subtilosin and rhizocticins | 9 |
| *B. subtilis* BBG100 | ATCC 6633 derivative overproducing mycosubtilin, Nm[r] | This study |
| *Erwinia chrysanthemi* 3937 | | 14 |
| *Micrococcus luteus* | | Lab stock |
| *Pseudomonas aeruginosa* 7NSK2 | | 12 |
| Fungi | | |
| *Botrytis cinerea* | Wild-type | Lab stock |
| *Fusarium oxysporum* | Wild-type | Lab stock |
| *Pythium aphanidermatum* | Wild-type | 26 |
| Yeasts | | |
| *Pichia pastoris* | Wild-type | Lab stock |
| *Saccharomyces cerevisiae* | Wild-type | Lab stock |
| Plasmids | | |
| pUC19 | Cloning vector, Ap[r] | Biolabs, Beverly, MA |
| pBG101 | 0.5kb *fenF* PCR fragment inserted in pUC19, Ap[r] | This study |
| pBG102 | 0.7kb *pbp* PCR fragment inserted in pUC19, Ap[r] | This study |
| pBG103 | 0.5kb *Sal*I – *Eco*RI *fenF* fragment from pBG101 inserted in pBG102, Ap[r] | This study |
| pBEST501 | pGEM4 carrying the $P_{repU}$ promoter and *neo* gene from pUB110, Nm[r] | 15 |
| pBG106 | $P_{repU}$-*neo* fragment inserted in pBG103, Ap[r] Nm[r] | This study |

[a] Ap[r], resistance to ampicillin; Nm[r], resistance to neomycin.

TABLE 2. Biomass and lipopeptide production by the wild-type 6633 strain and its BBG100 derivative after 72 h of growth

|  | Biomass g. l$^{-1}$ | Lipopeptide production (mg l$^{-1}$)[a] | |
| --- | --- | --- | --- |
|  |  | Mycosubtilin | Surfactin |
| Wild-type in flask | 3.23 (SD = 0.13) | 17 (SD = 0.5) | 15 (SD = 4.1) |
| BBG100 in flask | 3.19 (SD = 0.24) | 203 (SD = 12.6) | 10 (SD = 3.4) |
| Wild-type in bioreactor | 3.25 (SD = 0.35) | 4.35 (SD = 5.1) | 1.15 (SD = 1.2) |
| BBG100 in bioreactor | 4.45 (SD = 1.4) | 66 (SD = 0.7) | 3.35 (SD = 4.03) |

[a] values are mean data from 2 experiments
SD : Standard deviation

TABLE 3. Calculated mass values of $M+H^+$, $M+Na^+$ and $M+K^+$ ions corresponding to identified homologues of surfactins and mycosubtilins in culture extracts from *B. subtilis* ATCC 6633 and its overproducing derivative

| Lipopeptide | $M+H^+$ | $M+Na^+$ | $M+K^+$ |
|---|---|---|---|
| Surfactin $C_{13}$ | 1008.66 | 1030.64 | 1046.61 |
| Surfactin $C_{14}$ | 1022.67 | 1044.66 | 1060.63 |
| Surfactin $C_{15}$ | 1036.69 | 1058.67 | 1074.65 |
| Mycosubtilin $C_{15}$ | 1057.57 | 1079.55 | 1095.52 |
| Mycosubtilin $C_{16}$ | 1071.58 | 1093.56 | 1109.54 |
| Mycosubtilin $C_{17}$ | 1085.6 | 1107.58 | 1123.55 |

TABLE 4. Growth inhibition activities of supernatants obtained from growth of the ATCC 6633 wild-type strain and its BBG100 derivative

| Strain | Antagonistic activity [a] | | | |
|---|---|---|---|---|
| | Wild-type | | Strain BBG100 | |
| | Supernatant | Protease treated | Supernatant | Protease treated |
| *E. chrysanthemi* | - | - | - | - |
| *E. coli* | - | - | - | - |
| *P. aeruginosa* | - | - | - | - |
| *M. luteus* | +++ | - | +++ | - |
| *B. cinerea* | +/- | +/- | +++ | +++ |
| *F. oxysporum* | +/- | +/- | ++ | ++ |
| *P. aphanidermatum* | +/- | - | ++ | + |
| *P. pastoris* | - | - | ++ | ++ |
| *S. cerevisiae* | +/- | +/- | +++ | +++ |

[a] Intensity of antagonistic activity was rated on the basis of the size of growth inhibition zones from the wells in which supernatant samples were deposited to the edge of the spreading fungal mycelium or cell colony, - : 0 mm, +/- : 1 – 4 mm, + : 5 – 7 mm, ++ : 8 – 9 mm, +++ : 10 mm or more. ND, Not done.

TABLE 5. Effect of strain 6633 and of its overproducing derivative on the reduction of damping-off of tomato plants caused by *P. aphanidermatum*[a]

| Treatment | | Seedling emergence (%) | | | |
|---|---|---|---|---|---|
| Pathogen | Bacterium | Exp. 1 | Exp. 2 | Exp. 3 | Exp. 4 |
| - | None | ND [b] | 96 | 95 | ND |
| - | 6633 | ND | 90 | 92 | ND |
| - | BBG100 | ND | 93 | 88 | ND |
| + | None | 38 | 48 | 59 | 8 |
| + | 6633 | 31 | 25 | 42 | 6 |
| + | BBG100 | 53 | 59 | 69 | 34 |

[a] Two hundred seeds were used for each treatment in every experiment and the number of healthy plantlets was counted 12 days after planting.

[b] ND, Not done.

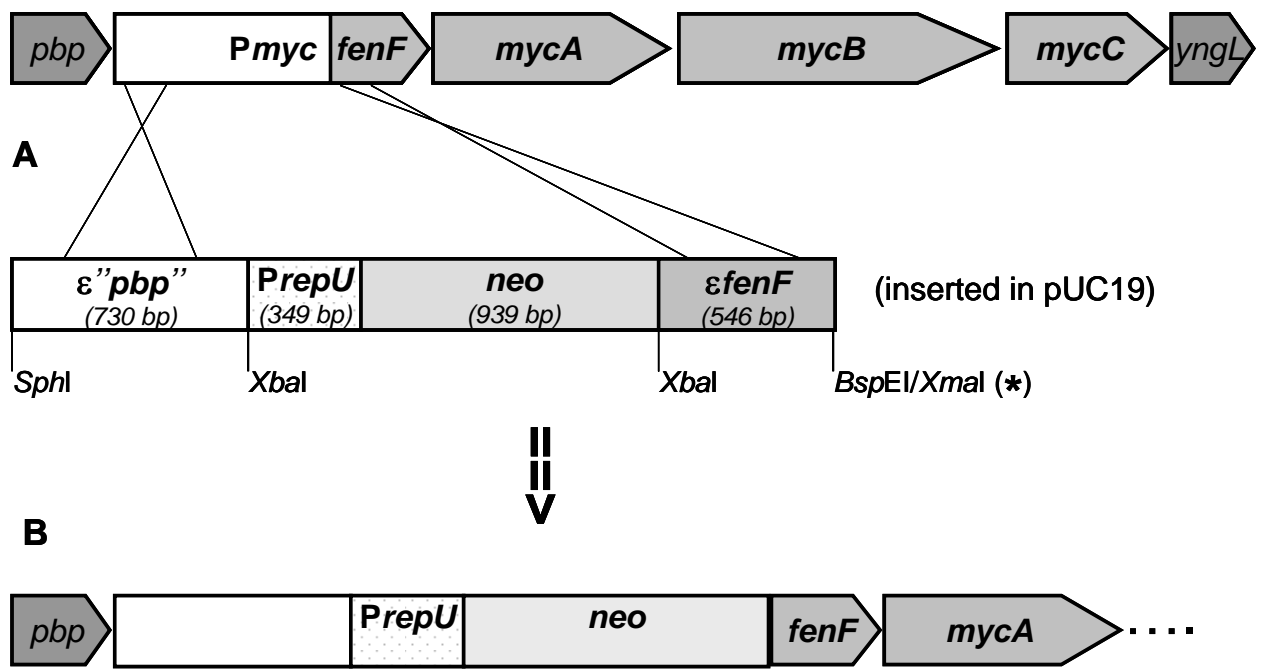

FIG. 1

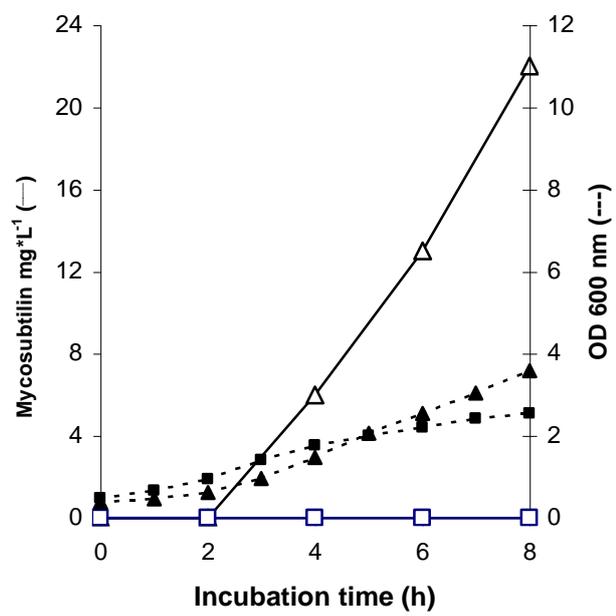

FIG. 2

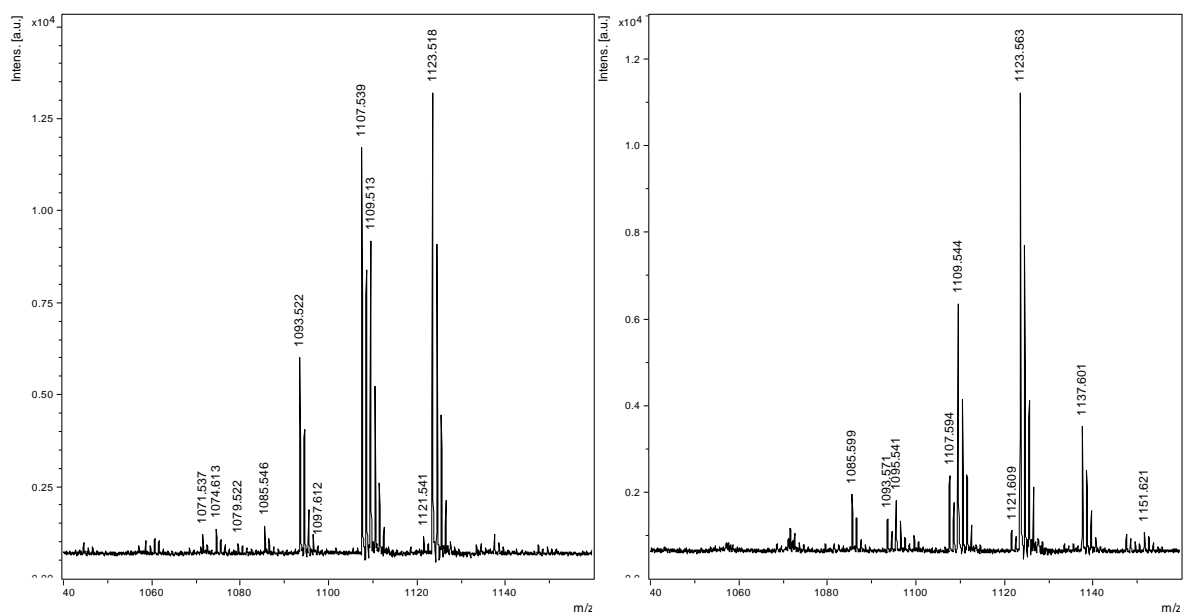

FIG. 3

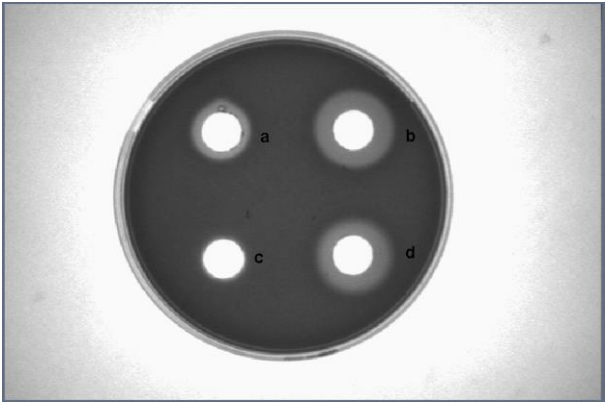

FIG.4

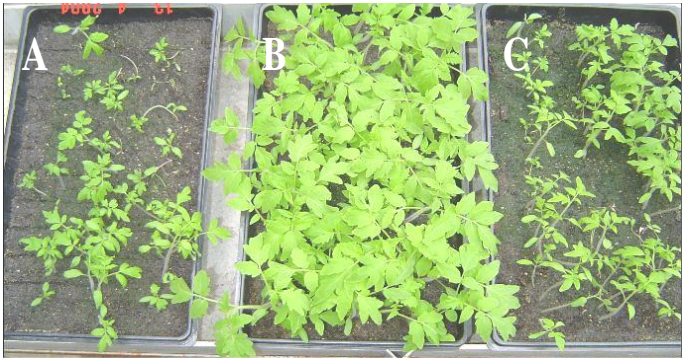

FIG. 5